\documentclass[twocolumn,showpacs,preprintnumbers,superscriptaddress,amsmath,amssymb]{revtex4}
\usepackage{graphicx}
\usepackage{dcolumn}
\usepackage{bm}
\usepackage{amssymb}
\usepackage{amssymb}
\def\>{\rangle}

\begin{document}

\title{Protecting entanglement in superconducting qubits}

\author{Jing Zhang}\email{jing-zhang@mail.tsinghua.edu.cn}
\affiliation{Advanced Science Institute, The Institute of Physical
and Chemical Research (RIKEN), Wako-shi, Saitama 351-0198, Japan}
\affiliation{Department of Automation, Tsinghua University,
Beijing 100084, P. R. China}
\author{Yu-xi Liu}
\affiliation{Advanced Science Institute, The Institute of Physical
and Chemical Research (RIKEN), Wako-shi, Saitama 351-0198,
Japan}\affiliation{CREST, Japan Science and Technology Agency
(JST), Kawaguchi, Saitama 332-0012, Japan}
\author{Chun-Wen Li}
\affiliation{Department of Automation, Tsinghua University,
Beijing 100084, P. R. China}
\author{Tzyh-Jong Tarn}
\affiliation{Department of Electrical and Systems Engineering,
Washington University, St. Louis, MO 63130, USA}
\author{Franco Nori}
\affiliation{Advanced Science Institute, The Institute of Physical
and Chemical Research (RIKEN), Wako-shi, Saitama 351-0198,
Japan}\affiliation{CREST, Japan Science and Technology Agency
(JST), Kawaguchi, Saitama 332-0012, Japan} \affiliation{Center for
Theoretical Physics, Physics Department, Center for the Study of
Complex Systems, The University of Michigan, Ann Arbor,
Michigan 48109-1040, USA}%

\date{\today}

\begin{abstract}
When a two-qubit system is initially maximally-entangled, two
independent decoherence channels, one per qubit, would greatly
reduce the entanglement of the two-qubit system when it reaches
its stationary state. We propose a method on how to minimize such
a loss of entanglement in open quantum systems. We find that the
quantum entanglement of general two-qubit systems with
controllable parameters can be protected by tuning both the
single-qubit parameters and the two-qubit coupling strengths.
Indeed, the maximum fidelity $F_{\rm max}$ between the stationary
entangled state, $\rho_{\infty}$, and the maximally-entangled
state, $\rho_m$, can be about $2/3\approx\max\{{\rm
tr}(\rho_{\infty}\rho_m)\}=F_{\rm max}$, corresponding to a
maximum stationary concurrence, $C_{\rm max}$, of about
$1/3\approx C(\rho_{\infty})=C_{\rm max}$. This is significant
because the quantum entanglement of the two-qubit system can be
protected, even for a long time. We apply our proposal to several
types of two-qubit superconducting circuits, and show how the
entanglement of these two-qubit circuits can be optimized by
varying experimentally-controllable parameters.

\end{abstract}

\pacs{85.25.-j, 03.67.Lx, 03.67.Mn}

 \maketitle

\section{Introduction}\label{s1}
Quantum information processing using superconducting qubits (see,
e.g.,~\cite{Makhlin,Wendin,You1,Clarkereview}) has made remarkable
advances in the past few years. One-qubit
and two-qubit quantum circuits (see, e.g.,
\cite{Pashkin,Yamamoto,Berkley,Izmalkov,McDermott,Majer,Clarke})
have been realized experimentally in superconducting systems. One
of the most important issues in quantum information processing is
how to couple two qubits, which has been widely studied
theoretically and experimentally in superconducting quantum
circuits (see,
e.g.,~\cite{Pashkin,Yamamoto,Berkley,Izmalkov,Majer,McDermott,Clarke,YuxiLiu,Paraoanu,Ashhab,Makhlin2,LC
osillator,RMigliore,Ploeg,Josephson junction,Nanomechanical
oscillator,Transmission line resonator1,Transmission line
resonator2,Blais2,Liu and Grajcar,Wei2}). To couple two qubits,
there are two types of approaches: (1) direct coupling and (2)
indirect coupling. Examples of qubit-qubit direct coupling
include, e.g., capacitively-coupled charge
qubits~\cite{Pashkin,Yamamoto} or inductively-coupled flux
qubits~\cite{Majer,Izmalkov,YuxiLiu}. Examples of indirect
coupling include, e.g., qubit-qubit coupling via a quantum $LC$
oscillator or an inductance~\cite{Makhlin2,LC osillator}, a
Josephson junction or an extra superconducting qubit acting as a
coupler~\cite{Ploeg,Josephson junction}, a nanomechanical
oscillator~\cite{Nanomechanical oscillator}, or a one or
three-dimensional transmission line resonator~\cite{Transmission
line resonator1,Transmission line resonator2,Blais2}. The main
merit of indirect coupling is that any two qubits can be
selectively coupled in a controllable way (see, e.g.,~\cite{Liu
and Grajcar}). By tuning some control parameters, one can
continuously adjust the coupling strengths between qubits, which
can be further used to switch between one-qubit and two-qubit
operations.

Although maximally-entangled states can be prepared either by
directly- or by indirectly-coupled two-qubit superconducting
quantum circuits~\cite{Wei2}, another equivalently important issue
is to {\it protect} these entangled states in an open environment.
Some entanglement-protection strategies have been proposed. For
instance, it has been recognized~\cite{Braun and Benatti,Nicolosi}
that a {\it common} dissipative environment, e.g., a heat bath,
would be helpful to protect the entanglement of two-qubit systems.
Specifically, a common dissipative environment would lead to a
collective decoherence channel which may induce a so-called
decoherence-free subspace~\cite{Duan} to protect special two-qubit
entangled states.

Even though a single collective decoherence channel could be used,
in principle, to protect two-qubit entangled states, it is very
difficult to obtain such a channel in experimental solid-state
systems (e.g., superconducting quantum circuits). {\it
Independent} decoherence channels usually lead to
disentanglement~\cite{Carvalho1}. This loss of entanglement cannot
be recovered by local operations and classical communications,
when the two qubits reach their stationary states. Even worse, a
mixture of the ``independent decoherence" and the ``collective
decoherence" channels may destroy the decoherence-free subspace
and lead to a failure of the entanglement protection. Thus, it is
very challenging to protect entanglement in the presence of such a
dissipative environment, with independent decoherence channels
acting on each qubit.

In this work, we propose a two-qubit entanglement-protection
strategy. We find that quantum entanglement can be partially
protected even when the two qubits interact {\it independently}
with their own separate environments. Our proposed strategy
requires that both the single-qubit oscillating frequencies and
the coupling strengths between qubits should be tunable.
Therefore, our proposal could be applied to several typical
superconducting quantum circuits, and other tunable solid-state
qubit systems.

This paper is organized as follows: in Sec.~\ref{s2} we present
our main results for general two-qubit systems. The entanglement
protection for directly- and indirectly-coupled superconducting
qubits via an inductive or a capacitive coupler is presented in
Sec.~\ref{s3} and Sec.~\ref{s4}. In Sec.~\ref{s5}, we study the
entanglement protection in superconducting qubit circuits
interacting with controllable squeezed modes in cavities or
resonators (e.g., circuit QED). Conclusions and discussions are
given in Sec.~\ref{s6}.

\section{General results}\label{s2}
We consider two coupled qubits with a general Hamiltonian
\begin{eqnarray}\label{General two-qubit Hamiltonian}
H_A&=&\mu_1\left[\exp(-i\theta_1)\sigma^{(1)}_{+}\sigma^{(2)}_{+}+\exp(i\theta_1)\sigma^{(1)}_{-}\sigma^{(2)}_{-}\right]\nonumber\\
&&+\mu_2\left[\exp(-i\theta_2)\sigma^{(1)}_{+}\sigma^{(2)}_{-}+\exp(i\theta_2)\sigma^{(1)}_{-}\sigma^{(2)}_{+}\right]\nonumber\\
&&+\sum_{j=1}^2\frac{\omega_{aj}}{2}\sigma^{(j)}_{z},
\end{eqnarray}
where the Planck constant $\hbar$ is assumed to be $1$. Here,
$\sigma^{(j)}_{\pm}=\sigma^{(j)}_{x}\pm i\sigma^{(j)}_{y}$, and
$\sigma^{(j)}_{x},\,\sigma^{(j)}_{y},\,\sigma^{(j)}_{z}$ are the
ladder and Pauli operators of the $j$-th qubit. The frequency of
the $j$-th qubit is denoted by $\omega_{aj}$. The real
coefficients $\mu_1$ (with phase $\theta_1$) and $\mu_2$ (with
phase $\theta_2$) correspond to the $\sigma^{(1)}_{-}
\sigma^{(2)}_{-}$ and $\sigma^{(1)}_{-} \sigma^{(2)}_{+}$ coupling
strengths, which are assumed to be tunable parameters.

The qubits also interact with uncontrollable degrees of freedom in
the environment. If the two qubits interact independently with
their own environments, then, under the Born-Markov
approximation~\cite{Walls}, we can obtain the following master
equation:
\begin{equation}\label{General master equation}
\dot{\rho}=-i[H_A,\rho]+\sum_{j=1}^2\Gamma_1\mathcal{D}[\sigma^{(j)}_{-}]\rho+\sum_{j=1}^2
2\Gamma_{\phi}\mathcal{D}[\sigma^{(j)}_{z}]\rho,
\end{equation}
where the super-operator $\mathcal{D}[L]\rho$ is defined as:
\begin{eqnarray*}
\mathcal{D}[L]\rho=L\rho
L^{\dagger}-\frac{1}{2}L^{\dagger}L\rho-\frac{1}{2}\rho
L^{\dagger}L,
\end{eqnarray*}
and $\Gamma_1,\,\Gamma_{\phi}$ represent the relaxation and pure
dephasing rates for each qubit, respectively. In order to simplify
our discussions, it is assumed that the two qubits have the same
relaxation and pure dephasing rates.

Below, we will use the concurrence $C(\rho)$:
\begin{equation}\label{Concurrence}
C(\rho)=\max\{\lambda_1-\lambda_2-\lambda_3-\lambda_4,0\},
\end{equation}
to quantify the quantum entanglement between the two qubits (see
Ref.~\cite{Wootters}), where the $\lambda_i$'s are the square
roots of the eigenvalues, in decreasing order, of the matrix
\begin{eqnarray*}
M=\rho(\sigma^{(1)}_{y}\sigma^{(2)}_{y})\rho^*(\sigma^{(1)}_{y}\sigma^{(2)}_{y}),
\end{eqnarray*}
and $\rho^*$ is the complex conjugate of the density matrix
$\rho$.

If the two qubits do not interact with each other, i.e.,
$\mu_1=\mu_2=0$ in Eq. (\ref{General master equation}), the
stationary state of the two qubits should be the ground state
$\rho^u_{\infty}=|00\rangle\langle00|$ (see,
e.g.,~\cite{Carvalho1}), where the superscript ``{\it u}" refers
to ``uncontrolled" qubit system. Since
$\rho_{\infty}^u=|00\rangle\langle00|$ is a separable state, the
entanglement between two qubits is completely {\it lost} even if
they are initially prepared in a maximally-entangled state.
However, the following discussions show that {\it the entanglement
can be partially protected by tuning the interaction strength
$\mu_1$ and the single-qubit frequencies $\omega_{aj}$ in
Eq.~(\ref{General two-qubit Hamiltonian}).}

\subsection{Strong interaction regime: $\mu_i\backsim10^{-1}(\omega_{a1}+\omega_{a2})$}\label{s21}

Let us now study the regime where the coupling strengths $\mu_1$
and $\mu_2$ are about one-order of magnitude smaller than the sum
of the two single-qubit frequencies:
$\Omega=\omega_{a1}+\omega_{a2}$, and the phases
$\theta_1,\,\theta_2$ are both time-independent parameters. For
such systems, we have the following results:

The solution $\rho(t)$ of Eq.~(\ref{General master equation})
tends to a stationary state
\begin{eqnarray}\label{Stationary entangled state for strong interaction regime}
\rho_{\infty}=p\rho_m+(1-p)\rho_s
\end{eqnarray}
as a convex combination of a maximally-entangled state $\rho_m$:
\begin{eqnarray*}
\rho_m&=&\frac{1}{2}\left(|00\rangle+e^{i(\theta_1-\phi)}|11\rangle\right)\left(\langle00|+e^{-i(\theta_1-\phi)}\langle11|\right)\\
&=&\frac{1}{2}\left(%
\begin{array}{cccc}
  1 &  &  & \exp[-i(\theta_1-\phi)] \\
   & 0 &  &  \\
   &  & 0 &  \\
  \exp[i(\theta_1-\phi)] &  &  & 1 \\
\end{array}%
\right)
\end{eqnarray*}
and a diagonal separable state $\rho_s$:
\begin{eqnarray*}
\rho_s=\left(%
\begin{array}{cccc}
  1-3\beta &  &  &  \\
   & \beta &  &  \\
   &  & \beta &  \\
   &  &  & \beta \\
\end{array}%
\right),
\end{eqnarray*}
where
\begin{eqnarray*}
\beta=\frac{1}{8}\left(1-\sqrt{1-\frac{8\Gamma_2}{\Gamma_1}p^2}\right).
\end{eqnarray*}
The subscript ``$m$" is an abbreviation of ``maximally-entangled",
and the subscript ``$s$" refers to ``separable".
$|0\rangle,\,|1\rangle$ are the two eigenstates of a single qubit.
The parameters $p,\,\phi$ can be expressed as:
\begin{eqnarray}\label{p and phi for strong interaction regime}
p&=&\frac{\sqrt{\Omega^2+64\Gamma_2^2}/8\mu_1}{2\Gamma_2/\Gamma_1+(\Omega^2+64\Gamma_2^2)/64\mu_1^2},\nonumber\\
\phi&=&\arctan\left(-\,\frac{8\Gamma_2}{\Omega}\right).
\end{eqnarray}
$\Gamma_2$ in Eq.~(\ref{p and phi for strong interaction regime})
is the dephasing rate that is defined as:
$\Gamma_2=\Gamma_1/2+\Gamma_{\phi}$.

The concurrence $C$ of the stationary state $\rho_{\infty}$
(hereafter called the stationary concurrence) and the fidelity $F$
between $\rho_{\infty}$ and the maximally-entangled state $\rho_m$
(hereafter called the stationary fidelity) are given by:
\begin{eqnarray}\label{Stationary concurrence and fidelity for strong interaction regime}
C(\rho_{\infty})&=&\max\left\{\frac{8\mu_1\sqrt{\Omega^2+64\Gamma_2^2}-64\mu_1^2\Gamma_2/\Gamma_1}{128\mu_1^2\Gamma_2/\Gamma_1+(\Omega^2+64\Gamma_2^2)},\,0\right\},\nonumber\\
F(\rho_{\infty})&=&{\rm
tr}(\rho_m\rho_{\infty})\nonumber\\
&=&\frac{4\mu_1\sqrt{\Omega^2+64\Gamma_2^2}-32\mu_1^2\Gamma_2/\Gamma_1}{128\mu_1^2\Gamma_2/\Gamma_1+(\Omega^2+64\Gamma_2^2)}+\frac{1}{2}.
\end{eqnarray}
The derivation of Eq.~(\ref{Stationary concurrence and fidelity
for strong interaction regime}) is given in
Appendix~\ref{Derivation of the maximum concurrence and fidelity}.

Our calculations show that the stationary concurrence
$C(\rho_{\infty})$ and fidelity $F(\rho_{\infty})$ are not
affected by the interaction strength $\mu_2$. Indeed, $\mu_2$
induces a coherent superposition of the two eigen-states
$|01\rangle$ and $|10\rangle$ (see Fig.~\ref{Fig of the energy
transitions between four eigen-states of the two-qubit systems}).
These two states always decay to the two-qubit ground state
$|00\rangle$, when subject to independent relaxation and dephasing
channels. Thus, $\mu_2$ does not affect the stationary state
$\rho_{\infty}$. However, the interaction strength $\mu_1$ induces
a coherent superposition of the two eigen-states $|00\rangle$ and
$|11\rangle$. Of course, $|00\rangle$ is already in the ground
state, while the state $|11\rangle$ can be partially recovered
from $|00\rangle$ by the coherent superposition caused by $\mu_1$.
Therefore, the stationary concurrence $C(\rho_{\infty})$ and
fidelity $F(\rho_{\infty})$ only depend on $\mu_1$.
\begin{figure}
\includegraphics[bb=146 575 388 763, width=8 cm, clip]{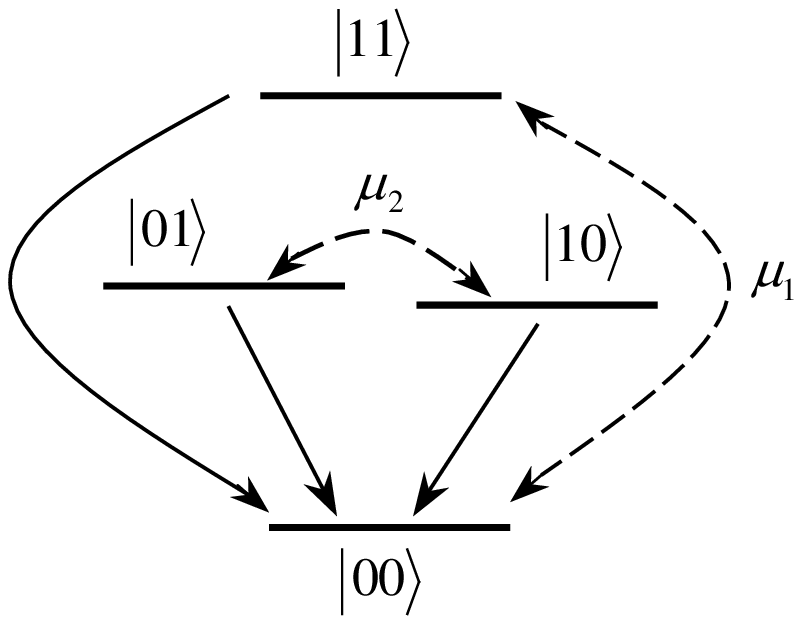}
\caption{ Schematic diagram of the energy transitions between the
four eigen-states: $|00\rangle$, $|11\rangle$, $|10\rangle$, and
$|01\rangle$. The solid arrows denote the decays caused by
independent relaxation and dephasing channels. The dashed arrows
represent the coherent superpositions caused by the interactions
between qubits (with interaction strengths $\mu_1$ and $\mu_2$,
respectively).}\label{Fig of the energy transitions between four
eigen-states of the two-qubit systems}
\end{figure}

Equation (\ref{Stationary concurrence and fidelity for strong
interaction regime}) shows that the maximum concurrence $C_{{\rm
max}}$ and the maximum fidelity $F_{{\rm max}}$, where
\begin{eqnarray}\label{Optimal
concurrence and fidelity for strong interaction regime}
C_{{\rm max}}&=&\frac{1}{4}\left(\sqrt{\frac{2\Gamma_1}{\Gamma_2}+1}-1\right),\nonumber\\
F_{{\rm
max}}&=&\frac{1}{8}\left(\sqrt{\frac{2\Gamma_1}{\Gamma_2}+1}-1\right)+\frac{1}{2},
\end{eqnarray}
can be obtained when the parameters $\mu_1$ and $\Omega$ satisfy
\begin{equation}\label{Optimal condition for strong interaction regime}
\mu_1=\frac{\Gamma_1}{8}\times\frac{\sqrt{\Omega^2+64\Gamma_2^2}}{\sqrt{2\Gamma_1\Gamma_2+\Gamma_2^2}+\Gamma_2}.
\end{equation}
The maximum concurrence $C_{\rm max}$ and fidelity $F_{\rm max}$,
given in Eq.~(\ref{Optimal concurrence and fidelity for strong
interaction regime}), are plotted in Fig.~\ref{Fig of the Copt and
Fopt}. This clearly shows that the concurrence $C_{\rm max}$ and
fidelity $F_{\rm max}$ in Eq.~(\ref{Optimal concurrence and
fidelity for strong interaction regime}) increase when the ratio
\begin{equation}\label{Gamma1 divided by Gamma2}
\frac{\Gamma_1}{\Gamma_2}=\frac{\Gamma_1}{\Gamma_1/2+\Gamma_{\phi}}
\end{equation}
increases, and the highest concurrence and fidelity
\begin{eqnarray}
\label{eq:9} C_{{\rm max}}&\rightarrow&\frac{\sqrt{5}-1}{4}\approx 0.31,\\
\label{eq:10} F_{{\rm
max}}&\rightarrow&\frac{\sqrt{5}+3}{8}\approx 0.65
\end{eqnarray}
can be obtained when $ \Gamma_2\rightarrow\Gamma_1/2$, i.e.,
$\Gamma_{\phi}\rightarrow0$.

\begin{figure}
\centerline{\includegraphics[width=2.5in,height=2.1in]{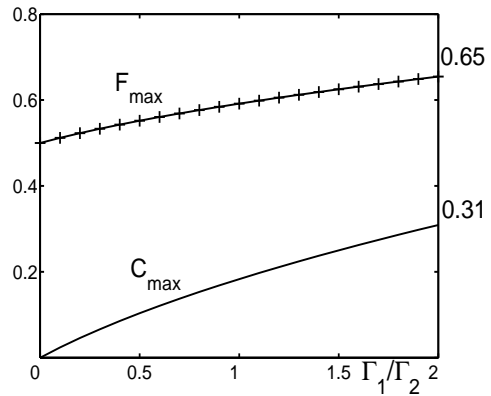}}
\caption{ Maximum concurrence $C_{{\rm max}}$ and maximum fidelity
$F_{{\rm max}}$, versus the ratio $\Gamma_1/\Gamma_2$ of the
relaxation rates, as given in Eq.~(\ref{Optimal concurrence and
fidelity for strong interaction regime}). Note that $C_{\rm
max}\rightarrow$ 0.31 and $F_{\rm max}\rightarrow$ 0.65 when
$\Gamma_1/\Gamma_2\rightarrow2$.}\label{Fig of the Copt and Fopt}
\end{figure}

The obtained maximum concurrence $C_{\rm max}$ may not be high
enough to be used for demanding tasks in quantum information
processing. However, there are two possible ways to increase the
stationary entanglement. First, the ratio $\Gamma_1/\Gamma_2$ in
Eq.~(\ref{Gamma1 divided by Gamma2}) can approach the optimal
value $2$ for certain systems. For example, if the superconducting
charge or flux qubits are at their degenerate
points~\cite{Makhlin,Wendin,You1}, the optimal value $2$ might be
obtained. In this case, the optimal concurrence and fidelity can
be obtained, as shown in Eqs.~(\ref{eq:9}) and (\ref{eq:10}).

Second, Eq.~(\ref{Optimal concurrence and fidelity for strong
interaction regime}) shows that the maximum fidelity $F_{\rm max}$
between the stationary state $\rho_{\infty}$ and the
maximally-entangled state $\rho_m$ is always larger than $0.5$,
which makes it possible to introduce additional entanglement
purification process to increase the proportion of the
maximally-entangled state $\rho_m$. Even if the concurrence and
fidelity reach the optimal values, as in Eqs.~(\ref{eq:9}) and
(\ref{eq:10}), we can, in principle, further increase the
stationary entanglement by using purification strategies, e.g., as
in Refs.~\cite{Dur,Maruyama and Tanamoto}.

\subsection{Weak interaction regime: $\mu_i\ll 10^{-1}(\omega_{a1}+\omega_{a2})$}\label{s22}

The results obtained in subsection \ref{s21} cannot be efficiently
applied to the case when $\mu_1,\,\mu_2\ll 10^{-1}\Omega$ (e.g.,
when $\mu_1,\,\mu_2$ are two orders of magnitude smaller than
$\Omega$). In fact, the optimal condition (\ref{Optimal condition
for strong interaction regime}) shows that we can obtain the
maximum concurrence $C_{\rm max}$ and fidelity $F_{\rm max}$ only
when
\begin{eqnarray*}
\mu_1=\frac{\Gamma_1}{8}\times\frac{\sqrt{\Omega^2+64\Gamma_2^2}}{\sqrt{2\Gamma_1\Gamma_2+\Gamma_2^2}+\Gamma_2}\geq\frac{\Gamma_1}{8(\sqrt{5}+1)\Gamma_2}\Omega.
\end{eqnarray*}
Thus, if $\mu_1\ll 10^{-1}\Omega$, then, optimally, the ratio
$\Gamma_{1}/\Gamma_{2}$ should be very small. In this case, from
Eq.~(\ref{Stationary concurrence and fidelity for strong
interaction regime}), the obtained maximum stationary concurrence
$C_{\rm max}$ will be extremely small. Alternatively, in order to
avoid this problem, a time-dependent interaction between qubits
should be introduced:

If the phase $\theta_1$ in Eq.~(\ref{General two-qubit
Hamiltonian}) could be tuned to be
\begin{eqnarray*}
\theta_1=\Omega t+\phi_0=(\omega_{a1}+\omega_{a2})t+\phi_0,
\end{eqnarray*}
the obtained long-time state $\rho_{\infty}$ will be a
time-dependent state
\begin{eqnarray*}
\rho_{\infty}(t)=\frac{\mu_1\Gamma_1}{2\mu_1^2+\Gamma_1\Gamma_2}\tilde{\rho}_m(t)+\left(1-\frac{\mu_1\Gamma_1}{2\mu_1^2+\Gamma_1\Gamma_2}\right)\tilde{\rho}_s
\end{eqnarray*}
as a convex combination of a time-dependent maximally-entangled
state
\begin{eqnarray*}
\tilde{\rho}_m(t)=\frac{1}{2}\left(%
\begin{array}{cccc}
  1 &  &  & e^{-i\left(\Omega t+\phi_0-\frac{\pi}{2}\right)} \\
   & 0 &  &  \\
   &  & 0 &  \\
  e^{i\left(\Omega t+\phi_0-\frac{\pi}{2}\right)} &  &  & 1 \\
\end{array}%
\right)
\end{eqnarray*}
and a time-independent diagonal separable state $\tilde{\rho}_s$:
\begin{eqnarray*}
\tilde{\rho}_s=\left(%
\begin{array}{cccc}
  1-3\tilde{\beta} &  &  &  \\
   & \tilde{\beta} &  &  \\
   &  & \tilde{\beta} &  \\
   &  &  & \tilde{\beta} \\
\end{array}%
\right),
\end{eqnarray*}
where
\begin{eqnarray*}
\tilde{\beta}=\frac{1}{8}\left(1-\sqrt{1-\frac{8\Gamma_2}{\Gamma_1}\left(\frac{\mu_1\Gamma_1}{2\mu_1^2+\Gamma_1\Gamma_2}\right)^2}\right).
\end{eqnarray*}
The corresponding stationary concurrence and fidelity can now be
expressed as:
\begin{eqnarray}\label{Stationary concurrence and fidelity for weak interaction regime}
C(\rho_{\infty})&=&\max\left\{\frac{\mu_1(\Gamma_1-\mu_1)}{2\mu_1^2+\Gamma_2\Gamma_1},0\right\},\nonumber\\
F(\rho_{\infty})&=&\frac{\mu_1(\Gamma_1-\mu_1)}{4\mu_1^2+2\Gamma_1\Gamma_2}+\frac{1}{2}.
\end{eqnarray}

For the same reason discussed in subsection~\ref{s21}, the
stationary concurrence $C(\rho_{\infty})$ and fidelity
$F(\rho_{\infty})$ are not affected by the interaction strength
$\mu_2$. The proof of Eq.~(\ref{Stationary concurrence and
fidelity for weak interaction regime}) is similar to the proof of
Eq.~(\ref{Stationary concurrence and fidelity for strong
interaction regime}). From Eq.~(\ref{Stationary concurrence and
fidelity for weak interaction regime}), the maximum stationary
concurrence and fidelity
\begin{eqnarray}\label{Optimal concurrence and fidelity for weak interaction regime}
C_{{\rm max}}&=&\frac{1}{4}\left(\sqrt{\frac{2\Gamma_1}{\Gamma_2}+1}-1\right),\nonumber\\
F_{{\rm
max}}&=&\frac{1}{8}\left(\sqrt{\frac{2\Gamma_1}{\Gamma_2}+1}-1\right)+\frac{1}{2}
\end{eqnarray}
can be obtained when
\begin{equation}\label{Optimal condition for weak interaction regime}
\mu_1=\frac{\Gamma_1\Gamma_2}{\sqrt{2\Gamma_1\Gamma_2+\Gamma_2^2}+\Gamma_2}.
\end{equation}
A higher stationary concurrence and fidelity can be obtained, as
in the strong interaction regime, by increasing the ratio
$\Gamma_1/\Gamma_2$.

Below, we apply the above results to several superconducting
circuits and discuss how their parameters can be varied so that
the stationary concurrence and fidelity can be maxima.

\section{Direct coupling between superconducting qubits}\label{s3}

\subsection{Two capacitively-coupled charge qubits}\label{s31}
\begin{figure}
\includegraphics[bb=106 572 370 763, width=8 cm, clip]{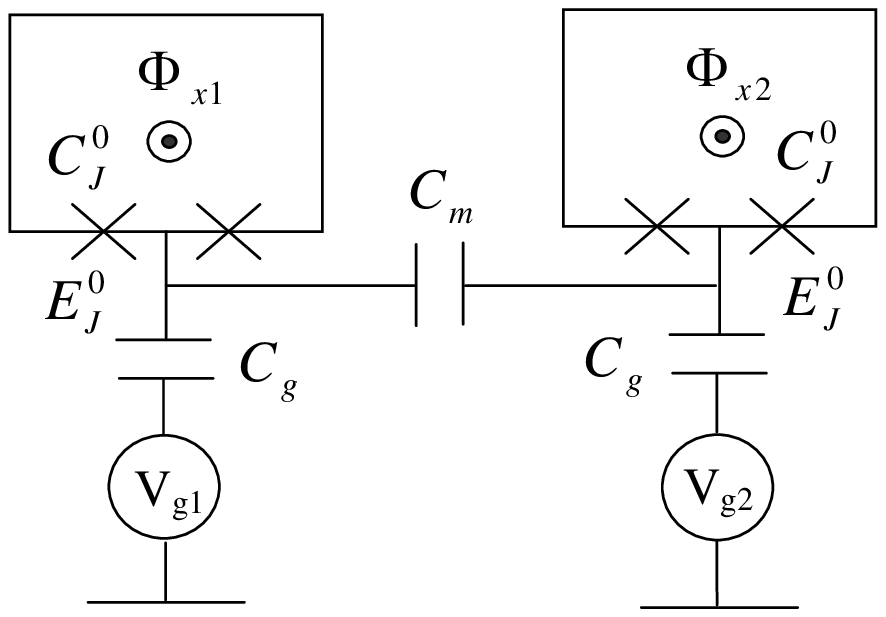}
\caption{ Schematic diagram of two capacitively coupled
CPBs.}\label{Fig of the capacitively coupled charge qubits}
\end{figure}

Let us first study the superconducting circuit shown in
Fig.~\ref{Fig of the capacitively coupled charge qubits}, where
two single Cooper pair boxes (CPBs) are connected via a small
capacitor~\cite{Yamamoto,Pashkin}. The Hamiltonian of the total
system can be
\begin{equation}\label{Original Hamiltonian of capacitively coupled charge qubits}
H_A=\sum_{j=1}^2
[4E_C(\hat{n}_j-n_{gj})^2-E_J(\Phi_{xj})\cos\hat{\phi}_j]+4J\hat{n}_1\hat{n}_2,
\end{equation}
where $\hat{\phi}_j$ is a phase operator denoting the phase drop
across the $j$-th CPB; $\hat{n}_j=-i\partial/(\partial
\hat{\phi}_j)$, which represents the number of Cooper pairs on the
island electrode,  is the conjugate operator of $\hat{\phi}_j$.
The reduced charge number $n_{gj}$, in units of the Cooper pairs
($2e$), can be given by $n_{gj}=-C_g V_{gj}/2e$, where the
parameters $C_g$ and $V_{gj}$ are the gate capacitance and gate
voltage of the $j$-th CPB. The Josephson energy $E_J(\Phi_{xj})$
of the $j$-th dc SQUID is
\begin{eqnarray*}
E_J(\Phi_{xj})=2E_J^0\cos\left(\pi\frac{\Phi_{xj}}{\Phi_0}\right),
\end{eqnarray*}
where $E_J^0$ represents the Josephson energy of a single
Josephson junction~\cite{Frank}; $\Phi_{xj}$ denotes the external
flux piercing the SQUID loop of the $j$-th CPB; and $\Phi_0$ is
the flux quantum. The coupling constant $J$ between two CPBs is
\begin{eqnarray*}
J=\frac{e^2 C_m}{(C_g+2C_J^0)^2-C_m^2},
\end{eqnarray*}
where $C_J^0$ and $C_m$ are the capacitance of a single Josephson
junction and the coupling capacitance between two CPBs.
$E_C=e^2/2(C_g+2C_J^0)$ is the single-electron charging energy of
a single CPB. For simplicity, we assume that $E_{C}$ and
$E_{J}^{0}$ are the same for the two CPBs.

Near $n_{gj}=0.5$, which is called the charge degenerate point,
the two energy levels of the $j$-th CPB corresponding to $n_j=0,1$
are close to each other and far separated from other high-energy
levels. In this case, a single CPB can be approximately considered
as a two-level system. In the charge basis, the Hamiltonian $H_A$
in Eq.~(\ref{Original Hamiltonian of capacitively coupled charge
qubits}) can be written as~\cite{Wendin}:
\begin{equation}\label{Hamiltonian of capacitively coupled charge qubits in the basis of charge states}
H_A=\sum_{j=1}^2
\left(-\frac{1}{2}E_C(n_{gj})\tilde{\sigma}^{(j)}_{z}-\frac{1}{2}E_J(\Phi_{xj})\tilde{\sigma}^{(j)}_{x}\right)+J\tilde{\sigma}^{(1)}_{z}\tilde{\sigma}^{(2)}_{z},
\end{equation}
where $E_C(n_{gj})=4E_C(1-2n_{gj})$ and the Pauli operators are
defined as:
\begin{eqnarray*}
&\tilde{\sigma}^{(j)}_{x}=|0\rangle_{j\,j}\langle1|+|1\rangle_{j\,j}\langle0|,&\\
&\tilde{\sigma}^{(j)}_{z}=|0\rangle_{j\,j}\langle0|-|1\rangle_{j\,j}\langle1|.&
\end{eqnarray*}
Here, $|0\rangle_j$ and $|1\rangle_j$ are the charge states with
the Cooper pair numbers $n_j=0,1$, respectively.

Rewriting Eq.~(\ref{Hamiltonian of capacitively coupled charge
qubits in the basis of charge states}) using the eigenstates of
the single-qubit Hamiltonian, we have
\begin{eqnarray*}
H_A=\sum_{j=1}^2\frac{\omega_{aj}}{2}\sigma^{(j)}_{z}+J\prod_{j=1}^2\left(\frac{E_{Jj}}{\omega_{aj}}\sigma^{(j)}_{x}-\frac{E_{Cj}}{\omega_{aj}}\sigma^{(j)}_{z}\right),
\end{eqnarray*}
with $\omega_{aj}=\left(E_{Cj}^2+E_{Jj}^2\right)^{1/2}$,
$E_{Cj}=E_C(n_{gj})$, and $E_{Jj}=E_J(\Phi_{xj})$. The new Pauli
operators $\sigma^{(j)}_{x}$ and $\sigma^{(j)}_{z}$ are defined by
the eigenstates $|+\rangle_j$ and $|-\rangle_j$ of the $j$-th
qubit as:
\begin{eqnarray*}
\sigma^{(j)}_{x}=|+\rangle_{j\,j}\langle-|+|-\rangle_{j\,j}\langle+|,\\
\sigma^{(j)}_{z}=|+\rangle_{j\,j}\langle+|-|-\rangle_{j\,j}\langle-|,
\end{eqnarray*}
where
\begin{eqnarray*}
|+\rangle_j=\cos\theta_j|0\rangle_j-\sin\theta_j|1\rangle_j,\\
|-\rangle_j=\sin\theta_j|0\rangle_j+\cos\theta_j|1\rangle_j.
\end{eqnarray*}
Here,
$\theta_j=\left[\arctan\left(-E_{Jj}/E_{Cj}\right)\right]/2$.

Now, the two-qubit state $\rho(t)$ evolves following the master
equation:
\begin{eqnarray*}
\dot{\rho}=-i[H_A,\rho]+\sum_{j=1}^2\Gamma_1\mathcal{D}[\sigma^{(j)}_{-}]\rho+\sum_{j=1}^2
2\Gamma_{\phi}\mathcal{D}[\sigma^{(j)}_{z}]\rho.
\end{eqnarray*}
Let us now assume that the two qubits are both in the charge
degenerate point $n_{gj}=0.5$ with $E_{Cj}=0$, so that the
dephasing effects can be minimized. In this case, we have
$\Gamma_{\phi}=0$, which means that
$\Gamma_2=\Gamma_1/2+\Gamma_{\phi}=\Gamma_1/2$. Further, from
$E_{Cj}=0$, we have:
\begin{eqnarray*}
H_A&=&\sum_{j=1}^2\frac{E_J(\Phi_{xj})}{2}\sigma^{(j)}_{z}+J\sigma^{(1)}_{x}\sigma^{(2)}_{x}\\
&=&\sum_{j=1}^2\frac{E_J(\Phi_{xj})}{2}\sigma^{(j)}_{z}+\frac{J}{4}(\sigma^{(1)}_{+}\sigma^{(2)}_{+}+\sigma^{(1)}_{-}\sigma^{(2)}_{-})\\
&&+\frac{J}{4}(\sigma^{(1)}_{+}\sigma^{(2)}_{-}+\sigma^{(1)}_{-}\sigma^{(2)}_{+}).
\end{eqnarray*}

In some experiments (e.g., ~\cite{Pashkin}), the coupling strength
$J$ is of the same order of
\begin{eqnarray*}
\Omega=\sum_{j=1}^2
2E_J^0\cos\left(\pi\frac{\Phi_{xj}}{\Phi_0}\right),
\end{eqnarray*}
where $J\approx4\,\,{\rm GHz}$, $E_J^0\approx10\,\,{\rm GHz}$ and
the decoherence rates~\cite{Astafiev} $\Gamma_1,\,\Gamma_2$ are of
the order of $10$--$100$ MHz $\ll J,\,\Omega$.

If we let $\Phi_{x1}=\Phi_{x2}=\Phi_x$ and substitute
\begin{eqnarray*}
\Omega=2 E_{J}(\Phi_x),\quad\mu_1=\frac{J}{4}
\end{eqnarray*}
into Eq.~(\ref{Optimal condition for strong interaction regime}),
then, in the limit $\Omega\gg\Gamma_1,\,\Gamma_2$, the maximum
concurrence $C_{{\rm max}}\approx 0.31$ and fidelity $F_{{\rm
max}}\approx 0.65$, as in Eqs.~(\ref{eq:9}) and (\ref{eq:10}), can
be obtained, if only
\begin{eqnarray*}
J
&\approx&\frac{4}{\sqrt{5}+1}E_J^0\cos\left(\pi\frac{\Phi_x}{\Phi_0}\right),
\end{eqnarray*}
or, equivalently,
\begin{eqnarray*}
\cos\left(\pi\frac{\Phi_x}{\Phi_0}\right)\approx\frac{\sqrt{5}+1}{4}\times\frac{J}{E_J^0}.
\end{eqnarray*}

\subsection{Two inductively-coupled flux qubits}\label{s32}
Let us now consider a superconducting circuit, as shown in
Fig.~\ref{Fig of the inductively coupled flux qubits}, where two
flux qubits are coupled through their mutual inductance. Here, we
modify the design used in the experimental device in
Ref.~\cite{Majer}. Namely, the small junction of each
three-junction flux qubit is replaced by a dc SQUID, from which we
can adjust the tunnelling amplitude between the left and right
wells (see Fig.~\ref{Fig of double-well potential of flux qubits})
of each single flux qubit.

\begin{figure}[h]
\includegraphics[bb=98 602 351 760, width=8 cm, clip]{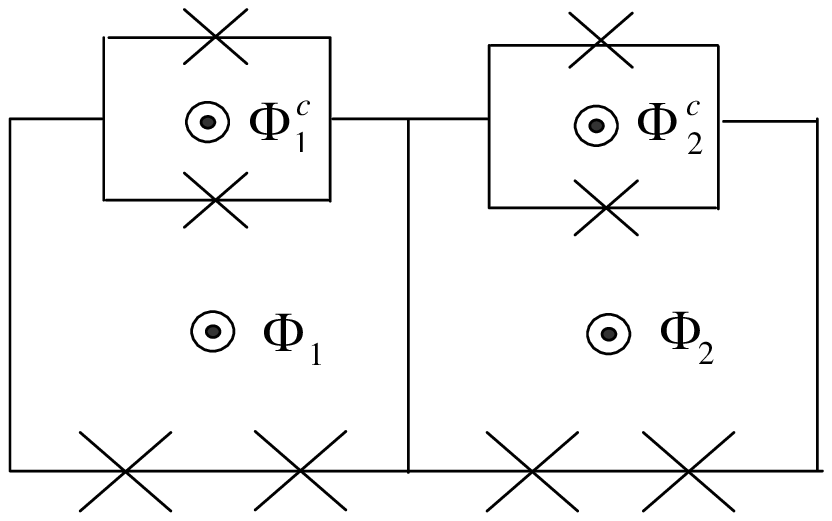}
\caption{ Schematic diagram of two flux qubits coupled via their
mutual inductance.}\label{Fig of the inductively coupled flux
qubits}
\end{figure}

Near the flux degenerate point $\Phi_j\approx\Phi_0/2$, with the
external flux $\Phi_j$ piercing the superconducting loop of the
$j$-th qubit, each flux qubit behaves as a two-level system. The
total Hamiltonian of the two-qubit system can be expressed
as~\cite{Majer}:
\begin{equation}\label{Hamiltonian for inductively coupled flux qubits}
H_A=-\frac{1}{2}\sum_{j=1}^2\left[\epsilon(\Phi_j)\tilde{\sigma}^{(j)}_{z}+\Delta(\Phi_{j}^{c})\tilde{\sigma}^{(j)}_{x}\right]+J\tilde{\sigma}^{(1)}_{z}\tilde{\sigma}^{(2)}_{z},
\end{equation}
where
\begin{eqnarray*}
&\tilde{\sigma}^{(j)}_{x}=|L_j\rangle\langle
R_j|+|R_j\rangle\langle
L_j|,&\\
&\tilde{\sigma}^{(j)}_{z}=|L_j\rangle\langle
L_j|-|R_j\rangle\langle R_j|,&
\end{eqnarray*}
and $|L_j\rangle,\,|R_j\rangle$ are the two lowest energy-states
in the left and right wells of the $j$-th flux qubit (see
Fig.~\ref{Fig of double-well potential of flux qubits}). The
parameter $\epsilon(\Phi_j)$ denotes the energy difference between
$|L_j\rangle$ and $|R_j\rangle$ which can be expressed as:
\begin{eqnarray*}
\epsilon(\Phi_j)=2I_{pj}\left(\Phi_j-\frac{1}{2}\Phi_0\right),
\end{eqnarray*}
where $I_{pj}$ is the circulating current in the loop of the
$j$-th qubit. The tunnelling amplitude $\Delta(\Phi_{j}^{c})$
between the two wells is tunable by varying the magnetic flux
$\Phi_{j}^{c}$ piercing the $j$-th dc-SQUID. In the limit
\begin{eqnarray*}
0<\frac{2\pi L_j I_c(\Phi_j^c)}{\Phi_0}-1\ll1,
\end{eqnarray*}
$\Delta(\Phi_j^c)$ can be approximately expressed
as~\cite{RMigliore}:
\begin{eqnarray*}
\Delta(\Phi_j^c)\approx\frac{3\Phi_0^2}{8\pi^2L_j}\left(1-\frac{\Phi_0}{2\pi
L_j I_c(\Phi_j^c)}\right)^2,
\end{eqnarray*}
where $L_j$ is the self-inductance of the superconducting loop of
the $j$-th flux qubit, and
\begin{eqnarray*}
I_c(\Phi_j^c)=2I_0\left|\cos\left(\frac{\pi\Phi_j^c}{\Phi_0}\right)\right|
\end{eqnarray*}
is the tunable critical current of the $j$-th dc-SQUID with $I_0$
being the maximum critical current. The coupling strength $J$
between the two flux qubits is
\begin{eqnarray*}
J=M I_{p1} I_{p2},
\end{eqnarray*}
with the mutual inductance $M$ between the two flux qubits.
\begin{figure}[h]
\includegraphics[bb=85 626 486 765, width=8 cm, clip]{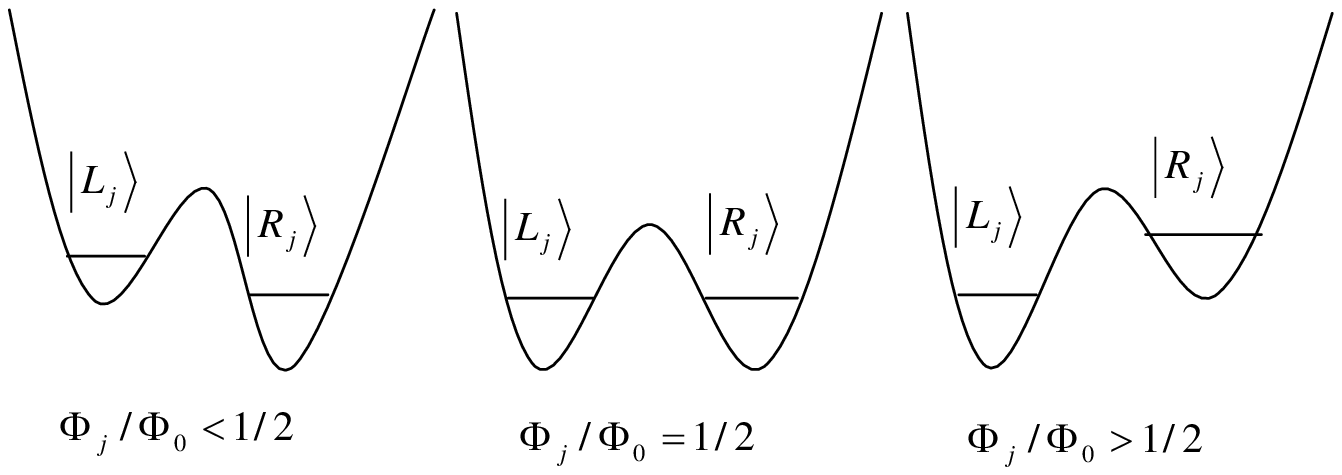}
\caption{ Schematic diagram of the double-well potential of the
$j$-th flux qubit with the two lowest-energy states for
$\Phi_j<\Phi_0/2$, $\Phi_j=\Phi_0/2$, and
$\Phi_j>\Phi_0/2$.}\label{Fig of double-well potential of flux
qubits}
\end{figure}

Let us now assume that $\Phi_j=\Phi_0/2$, then the two-qubit
Hamiltonian in Eq.~(\ref{Hamiltonian for inductively coupled flux
qubits}) can be further simplified to
\begin{eqnarray*}
H_A&=&\frac{1}{2}\sum_{j=1}^2\Delta(\Phi_j^c)\sigma^{(j)}_{z}+\frac{J}{4}(\sigma^{(1)}_{+}\sigma^{(2)}_{+}+\sigma^{(1)}_{-}\sigma^{(2)}_{-})\\
&&+\frac{J}{4}(\sigma^{(1)}_{+}\sigma^{(2)}_{-}+\sigma^{(1)}_{-}\sigma^{(2)}_{+}),
\end{eqnarray*}
where
\begin{eqnarray*}
&\sigma^{(j)}_{x}=|+\rangle_{j\,j}\langle-|+|-\rangle_{j\,j}\langle+|,&\\
&\sigma^{(j)}_{z}=|+\rangle_{j\,j}\langle+|-|-\rangle_{j\,j}\langle-|,&
\end{eqnarray*}
and
\begin{eqnarray*}
|+\rangle_j=\frac{\sqrt{2}}{2}\left\{|L_j\rangle-|R_j\rangle\right\},\\
|-\rangle_j=\frac{\sqrt{2}}{2}\left\{|L_j\rangle+|R_j\rangle\right\}.
\end{eqnarray*}
The decoherence process can be described by the master
equation~(\ref{General master equation}) under the Born-Markov
approximation. At the degenerate point
($\Phi_j=\Phi_0/2;\,j=1,2$), we have $\Gamma_{\phi}=0$, which
means that $\Gamma_2=(\Gamma_1/2)+\Gamma_{\phi}=\Gamma_1/2$.

Let us assume that $\Phi_1^c=\Phi_2^c=\Phi^c$ and substitute
\begin{eqnarray*}
\Omega=2\Delta(\Phi^c),\quad\mu_1=\frac{J}{4}
\end{eqnarray*}
into Eq.~(\ref{Optimal condition for strong interaction regime}),
then, in the limit $\Delta(\Phi^c),J\gg\Gamma_1,\,\Gamma_2$, the
maximum concurrence $C_{{\rm max}}\approx0.31$ and fidelity
$F_{{\rm max}}\approx0.65$ can be obtained when
\begin{eqnarray*}
\Delta(\Phi^c)\approx\frac{\sqrt{5}+1}{2}J.
\end{eqnarray*}

In experiments~\cite{Majer}, $\Delta(\Phi^c)$ and $J$ are of the
same order ($\sim 1$ GHz) that are far larger than the decoherence
rates $\Gamma_1,\,\Gamma_2\sim 1$--$10$ MHz (see,
e.g.,~\cite{Bertet}). Thus, the above optimal conditions could be
realized in experiments.

\section{Tunable coupling between superconducting qubits: strong-interaction
regime}\label{s4}

There are two ways to tune the system parameters to achieve the
optimal condition (\ref{Optimal condition for strong interaction
regime}). One way is by {\it tuning the sum of the single-qubit
oscillating frequencies $\Omega$}, which was used in
Sec.~\ref{s3}. In this section, we study another way to achieve
the optimal condition (\ref{Optimal condition for strong
interaction regime}) by {\it tuning the coupling strength
$\mu_1$} between the two qubits. 

\subsection{Variable-coupling between two charge qubits}\label{s41}
Many strategies have been proposed to obtain a controllable
coupling between qubits (see, e.g., ~\cite{Makhlin2,LC
osillator,RMigliore,Ploeg,Josephson junction,Nanomechanical
oscillator,Transmission line resonator1,Transmission line
resonator2,Blais2,Liu and Grajcar}). Let us first study the
superconducting circuit shown in Fig.~\ref{Fig of two charge
qubits coupled via a LC oscillator}, where two CPBs are coupled
via an $LC$ oscillator. This strategy was first proposed in
Ref.~\cite{Makhlin2} and also investigated by other researchers
(e.g., in Ref.~\cite{You and Yamamoto}). In the charge degenerate
point, the two-qubit Hamiltonian in Refs.~\cite{Makhlin,Makhlin2}
is
\begin{equation}\label{Hamiltonian for two charge qubits coupled via a LC oscillator}
H_A=-\sum_{j=1}^2\frac{1}{2}E_J(\Phi_{xj})\tilde{\sigma}^{(j)}_{x}
-E_{\rm int}\;\tilde{\sigma}^{(1)}_{y}\tilde{\sigma}^{(2)}_{y},
\end{equation}
with $E_{\rm int}=E_J(\Phi_{x1})E_J(\Phi_{x2})/E_L$ and
$E_J(\Phi_{xj})=2E_J^0\cos\left(\pi\Phi_{xj}/\Phi_0\right)$.  Here
\begin{eqnarray*}
\tilde{\sigma}^{(j)}_{x}&=&|0\rangle_{j\,j}\langle1|+|1\rangle_{j\,j}\langle0|,\\
\tilde{\sigma}^{(j)}_{y}&=&-i|0\rangle_{j\,j}\langle1|+i|1\rangle_{j\,j}\langle0|,
\end{eqnarray*}
and $|0\rangle_j,\,|1\rangle_j$ are the two charge states of the
$j$-th CPB. The quantity $E_L$ in the expression above for the
coupling strength $E_{\rm int}$ can be written as:
\begin{eqnarray*}
E_L=\left(\frac{2C_J^0}{C_{qb}}\right)^2\frac{\Phi_0^2}{\pi^2 L},
\end{eqnarray*}
where $C_{qb}=2C_J^0C_g\left[2C_J^0+C_g\right]^{-1}$ is the
capacitance of a single CPB in the external circuit, and $L$ is
the inductance of the coupling current-biased inductor.

\begin{figure}[h]
\includegraphics[bb=107 553 426 762, width=8 cm, clip]{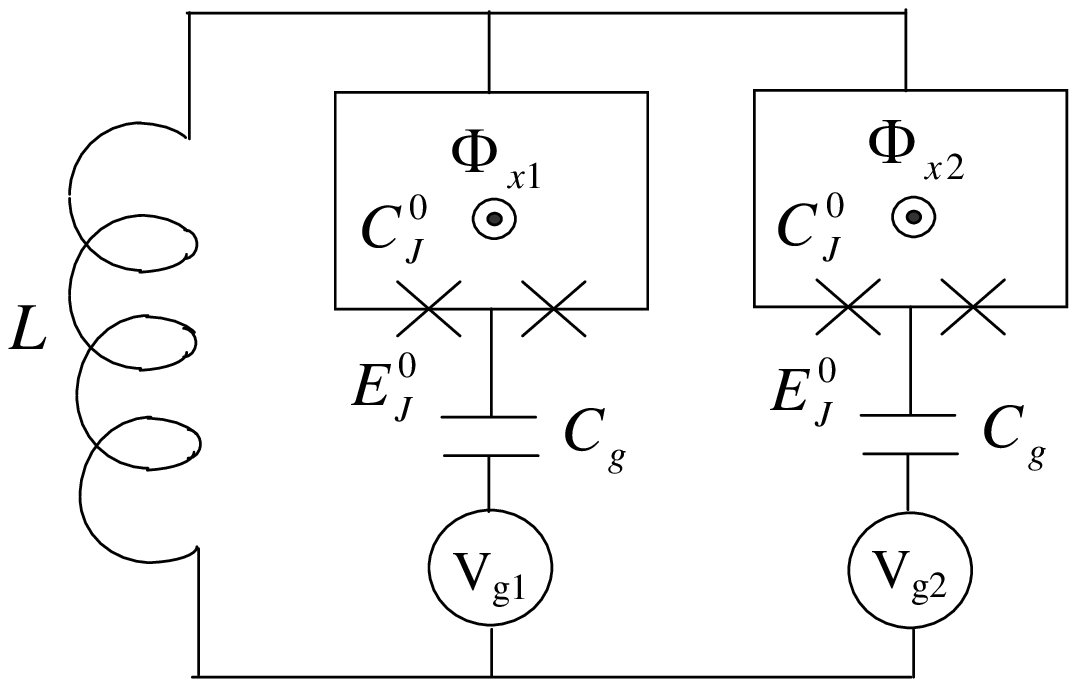}
\caption{ Schematic diagram of two charge qubits coupled via an
$LC$ oscillator.}\label{Fig of two charge qubits coupled via a LC
oscillator}
\end{figure}

Rewriting Eq.~(\ref{Hamiltonian for two charge qubits coupled via
a LC oscillator}) under the eigenstates of the single-qubit
Hamiltonian, we have:
\begin{eqnarray*}
H_A&=&\sum_{j=1}^2\frac{E_J(\Phi_{xj})}{2}\sigma^{(j)}_{z}+\frac{E_{\rm int}}{4}(\sigma^{(1)}_{+}\sigma^{(2)}_{+}+\sigma^{(1)}_{-}\sigma^{(2)}_{-})\\
&-&\frac{E_{\rm
int}}{4}(\sigma^{(1)}_{+}\sigma^{(2)}_{-}+\sigma^{(1)}_{-}\sigma^{(2)}_{+}),
\end{eqnarray*}
where
\begin{eqnarray*}
\sigma^{(j)}_{z}=|+\rangle_{j\,j}\langle+|-|-\rangle_{j\,j}\langle-|=-\tilde{\sigma}^{(j)}_{x},\\
\sigma^{(j)}_{y}=-i|+\rangle_{j\,j}\langle-|+i|-\rangle_{j\,j}\langle+|=\tilde{\sigma}^{(j)}_{y},
\end{eqnarray*}
and
\begin{eqnarray*}
|+\rangle_j=\frac{1}{\sqrt{2}}\left\{|0\rangle_j-|1\rangle_j\right\},\\
|-\rangle_j=\frac{1}{\sqrt{2}}\left\{|0\rangle_j+|1\rangle_j\right\}.
\end{eqnarray*}
Now, let $\Phi_{x1}=\Phi_{x2}=\Phi_x$. Since $\Gamma_2=\Gamma_1/2$
at the charge degenerate point, then, replacing $\Omega$ and
$\mu_1$ in Eq.~(\ref{Optimal condition for strong interaction
regime}) by $2E_J(\Phi_x)$ and $E_{\rm int}/4$, the maximum
concurrence $C_{{\rm max}}\approx 0.31$ and fidelity $F_{{\rm
max}}\approx 0.65$ can be obtained when
\begin{eqnarray*}
\cos\left(\pi\frac{\Phi_x}{\Phi_0}\right)\approx\frac{1}{\sqrt{5}+1}\times\frac{E_L}{E_J^0},
\end{eqnarray*}
when these conditions hold: $E_L,\,E_J^0\gg\Gamma_1,\,\Gamma_2$.

\subsection{Variable-coupling between two flux qubits}\label{s42}

Our strategy can also be applied to flux qubits with controllable
coupling. Here, let us consider a superconducting circuit design
as in Ref.~\cite{Ploeg} (see Fig.~\ref{Fig of two three-junction
flux qubits coupled via an auxiliary flux qubit}), where two
three-junction flux qubits (qubits $1$ and $2$) are coupled via an
auxiliary three-junction flux qubit (qubit $3$). This middle flux
qubit (qubit $3$), acting as a coupler, is connected to qubits $1$
and $2$ by sharing junctions $a$ and $b$ with the same Josephson
energy $E_J^0$, while junction $c$ is smaller than $a$ and $b$,
with Josephson energy $\alpha E_J^0,\,\alpha<1$. By adiabatically
eliminating the degrees of freedom of the auxiliary flux qubit
$3$, the total Hamiltonian of the flux qubits $1$ and $2$
becomes~\cite{Ploeg}
\begin{eqnarray*}
H_A&=&-\frac{1}{2}\sum_{j=1}^2[\epsilon(\Phi_j)\tilde{\sigma}^{(j)}_{z}+\Delta_j\tilde{\sigma}^{(j)}_{x}]+J(\Phi_3)\tilde{\sigma}^{(1)}_{z}\tilde{\sigma}^{(2)}_{z},
\end{eqnarray*}
where $\epsilon(\Phi_j)$, $\Delta_j,\,j=1,2$,
$\tilde{\sigma}_z^{(j)}$ and $\tilde{\sigma}_x^{(j)}$ have the
same meaning as in subsection \ref{s3}.B. When $\alpha\ll 1$, the
coupling strength $J(\Phi_3)$ between the two flux qubits $1$ and
$2$ becomes~\cite{Ploeg}:
\begin{equation}\label{Variable coupling strength between flux qubits}
J(\Phi_3)\approx\frac{\alpha I_{p1}
I_{p2}}{4e^2E_J^0}\cos\left(2\pi\frac{\Phi_3}{\Phi_0}\right)\equiv
J_0\cos\left(2\pi\frac{\Phi_3}{\Phi_0}\right).
\end{equation}
From Eq.~(\ref{Variable coupling strength between flux qubits}),
the coupling strength $J(\Phi_3)$ is tunable by varying the flux
$\Phi_3$ piercing the superconducting loop of the auxiliary qubit
$3$.

\begin{figure}[h]
\includegraphics[bb=95 596 377 765, width=8 cm, clip]{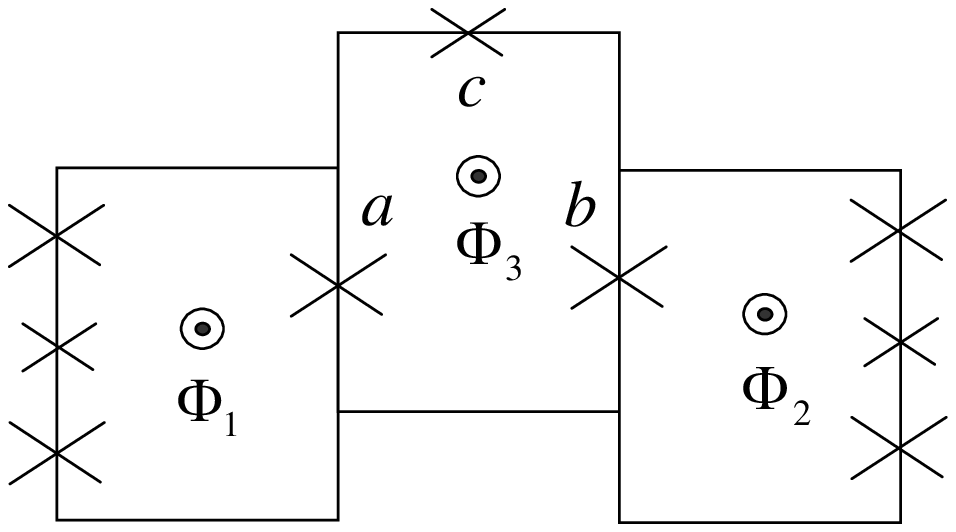}
\caption{ Schematic diagram of two three-junction flux qubits
coupled via an auxiliary flux qubit.}\label{Fig of two
three-junction flux qubits coupled via an auxiliary flux qubit}
\end{figure}

At the flux degenerate point, i.e., $\Phi_j=\Phi_0/2$ for both
qubits, and using the eigenstates of the single-qubit Hamiltonian,
we have:
\begin{eqnarray*}
H_A&=&\frac{1}{2}\sum_{j=1}^2\Delta_j\sigma^{(j)}_{z}+J(\Phi_3)\sigma^{(1)}_{x}\sigma^{(2)}_{x}\\
&=&\frac{1}{2}\sum_{j=1}^2\Delta_j\sigma^{(j)}_{z}+\frac{J(\Phi_3)}{4}(\sigma^{(1)}_{+}\sigma^{(2)}_{+}+\sigma^{(1)}_{-}\sigma^{(2)}_{-})\\
&&+\frac{J(\Phi_3)}{4}(\sigma^{(1)}_{+}\sigma^{(2)}_{-}+\sigma^{(1)}_{-}\sigma^{(2)}_{+}).
\end{eqnarray*}
Since the two flux qubits are at their flux degenerate points,
then, replacing $\Omega$ and $\mu_1$ in Eq.~(\ref{Optimal
condition for strong interaction regime}) by $\Delta_1+\Delta_2$
and $J(\Phi_3)/4$, the maximum stationary concurrence $C_{{\rm
max}}\approx0.31$ and fidelity $F_{{\rm max}}\approx0.65$ can be
obtained when $J(\Phi_3)\approx(\Delta_1+\Delta_2)/(\sqrt{5}+1)$,
i.e.,
\begin{eqnarray*}
\cos\left(2\pi\frac{\Phi_3}{\Phi_0}\right)\approx\frac{(\Delta_1+\Delta_2)}{(\sqrt{5}+1)J_0},
\end{eqnarray*}
when these conditions hold:
$\Delta_1,\,\Delta_2,\,J_0\gg\Gamma_1,\Gamma_2$.

From the experiment~\cite{Ploeg}, where $\Delta_1,\,\Delta_2$ and
$J_0$ are of the same order ($\sim 1$ GHz) and far larger than
$\Gamma_1,\,\Gamma_2$ ($\sim 1$--$10$ MHz), the above optimal
condition could be satisfied by varying the magnetic flux $\Phi_3$
through the middle superconducting loop.

\section{Tunable coupling between superconducting qubits: weak interaction
regime}\label{s5}

In this section, we study how to protect quantum entanglement in
superconducting circuits where two charge qubits are coupled to a
one-dimensional transmission line resonator. Since the interaction
strength ($10$--$100$ MHz) between the two charge qubits coupled
via the resonator is far smaller than the single-qubit oscillating
frequency ($5$--$15$ GHz), then we are now considering the
weak-interaction regime. From the analysis in subsection
\ref{s22}, in order to protect entanglement in this case, the
following time-dependent interaction Hamiltonian should be
introduced:
\begin{eqnarray*}
H_{{\rm int}}=\mu_1(e^{-i(\Omega
t+\phi_0)}\sigma^{(1)}_{+}\sigma^{(2)}_{+}+e^{i(\Omega
t+\phi_0)}\sigma^{(1)}_{-}\sigma^{(2)}_{-}).
\end{eqnarray*}

The main idea in this section is the following: borrowing
strategies that produce controllable squeezed fields in optical
cavities (see, e.g., \cite{Villas-Boas and Almeida}), an auxiliary
flux qubit circuit is introduced to squeeze the oscillating mode
in the resonator (see Fig.~\ref{Fig of the superconducting
circuit}). The auxiliary flux qubit circuit in fact acts like a
$\Delta$-shaped three-level atom which is further driven by a
classical field. By adiabatically eliminating the degrees of
freedom of the auxiliary flux qubit circuit, one can obtain a
controllable squeezed field in the resonator where the squeezed
coefficient is tunable by changing the coupling strength between
the classical driving field and the auxiliary flux qubit. With the
help of the controllable squeezed field in the resonator, one can
continuously adjust the stationary entanglement between the two
qubits.

\begin{figure}[h]
\includegraphics[bb=116 508 437 755, width=8 cm, clip]{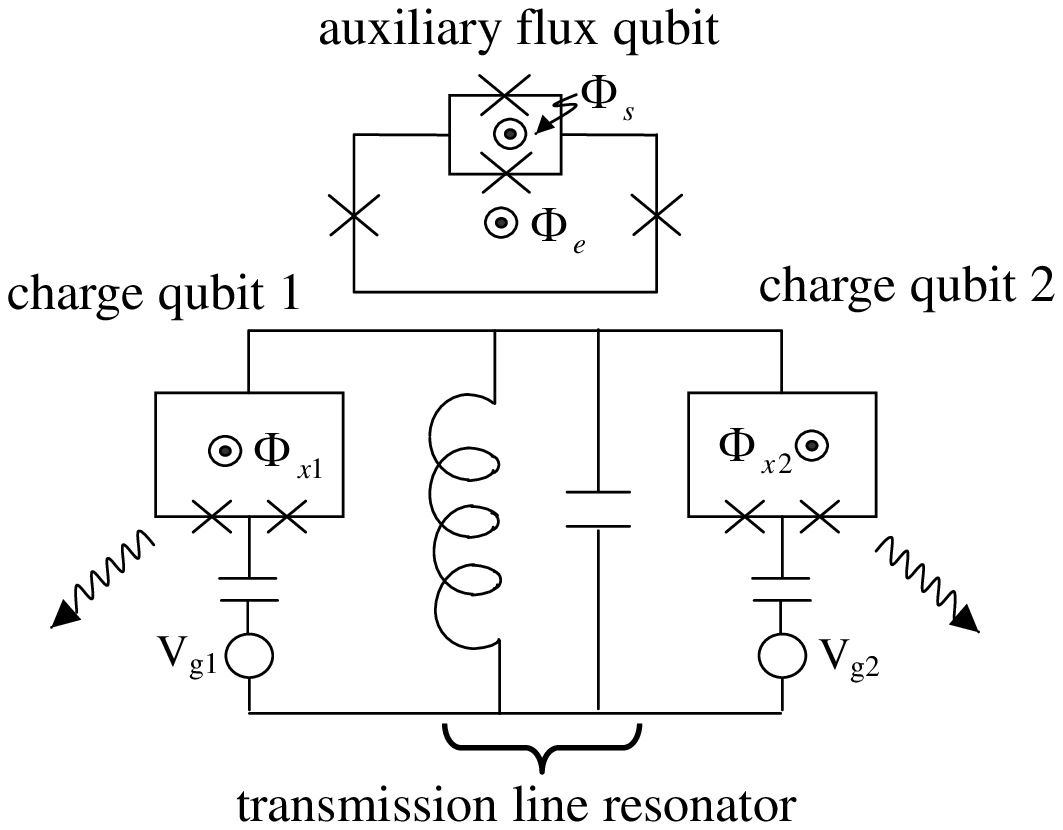}
\caption{ Schematic diagram of our proposal for protecting quantum
entanglement in two charge qubits coupled to a
resonator.}\label{Fig of the superconducting circuit}
\end{figure}

\subsection{Controllable squeezed electric field in a transmission line resonator}\label{s51}

We first show how to obtain a controllable squeezed electric
field~\cite{Hu,Hu2,Everitt,AMZagoskin} in the resonator by using a
theoretical proposal of realizing squeezed states in
cavities~\cite{Villas-Boas and Almeida}. The auxiliary flux qubit
circuit in our proposal (shown in Fig.~\ref{Fig of the
superconducting circuit}) acts as a three-level system with
$\Delta$-type transition~\cite{Liu3,You4}.

\begin{figure}[h]
\includegraphics[bb=86 523 506 765, width=8 cm, clip]{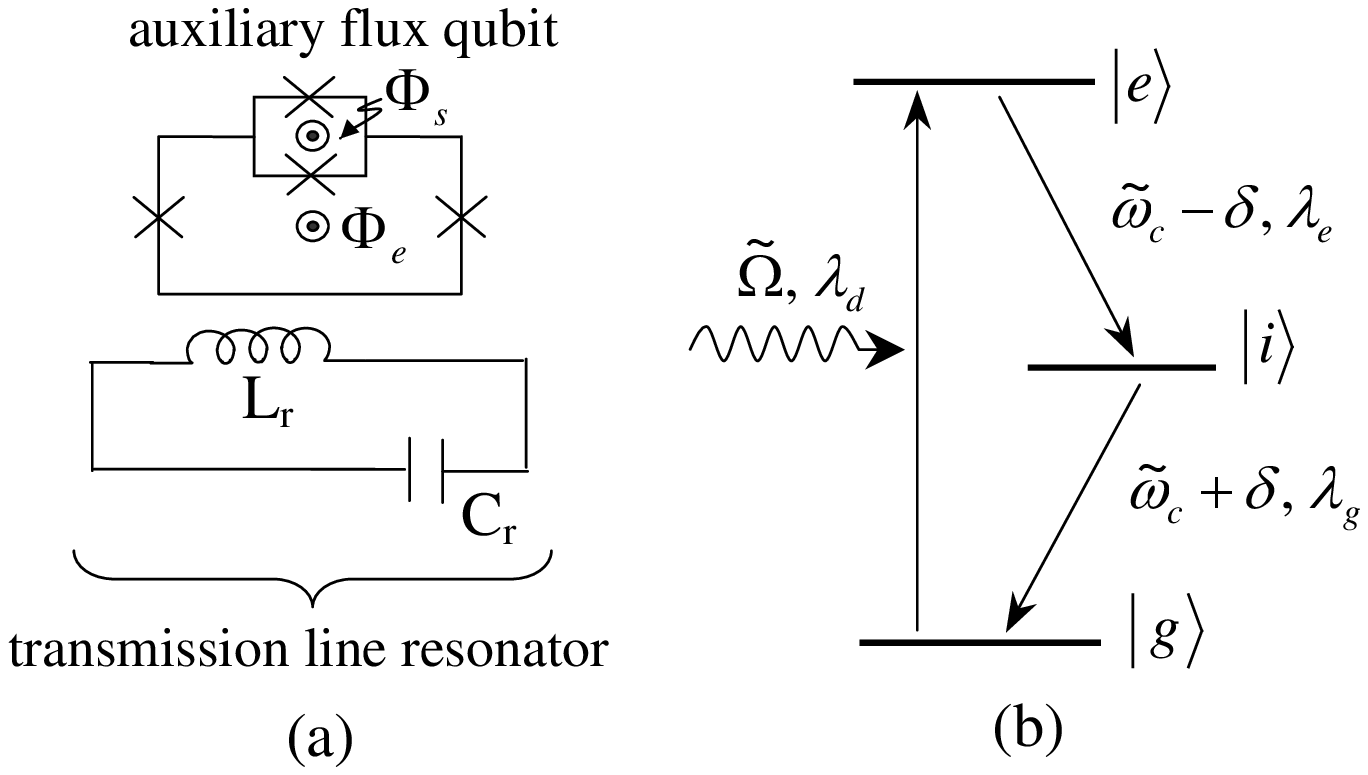}
\caption{ Schematic diagrams for realizing a controllable squeezed
electromagnetic field in a resonator coupled to an auxiliary flux
qubit. (a) The auxiliary flux qubit and the transmission line
resonator: the parameters in the auxiliary flux qubit are the same
as those in Ref.~\cite{You4}; by varying the flux $\Phi_s$
threading through the SQUID loop one can obtain a $\Delta$-shaped
three-level artificial atom. (b) Transition energy-level-diagram
of the $\Delta$-shaped three-level artificial atom.}\label{Fig of
squeezed state engineering technique}
\end{figure}

As depicted in Fig.~\ref{Fig of squeezed state engineering
technique}, we are now considering a three-level system with a
ground energy level $|g\rangle$, an intermediate energy level
$|i\rangle$, and an excited energy level $|e\rangle$. Here, the
transitions $|g\rangle\leftrightarrow|i\rangle$ and
$|e\rangle\leftrightarrow |i\rangle$ are coupled dispersively to
the quantized cavity mode in the resonator, with coupling
strengths $\lambda_g$ and $\lambda_e$. The transition
$|g\rangle\leftrightarrow|e\rangle$ is coupled dispersively to a
classical field with coupling strength $\lambda_d$ and frequency
$\tilde{\Omega}$. In the rotating-wave approximation, the total
Hamiltonian of the three-level artificial atom and the resonator
can be expressed as $H=H_0+V$, with
\begin{eqnarray}\label{Hamiltonian of the resonator and the auxiliary flux qubit}
H_0&=&\tilde{\omega}_c a^{\dagger}a-\tilde{\omega}_c
|g\rangle\langle g|+\delta|i\rangle\langle
i|+\tilde{\omega}_c|e\rangle\langle
e|,\nonumber\\
V&=&(\lambda_g a|i\rangle\langle g|+{\rm h.c.})+(\lambda_e
a|e\rangle\langle i|+{\rm h.c.})\nonumber\\
&&+(\lambda_d|e\rangle\langle g|e^{-i\Omega t}+{\rm h.c.}),
\end{eqnarray}
where ${\rm h.c.}$ means Hermitian conjugate; $\tilde{\omega}_c$
is the frequency of the resonator; $\delta$ is defined as a
detuning from the energy levels $|e\rangle$ and $|g\rangle$ to the
intermediate energy level $|i\rangle$.

Let us initially prepare the artificial atom in the intermediate
level $|i\rangle$. With the help of the dispersive-detuning
condition:
\begin{eqnarray*}
\delta\gg
|\lambda_g|,\,|\lambda_e|,\,|\tilde{\Omega}-2\tilde{\omega}_c|,
\end{eqnarray*}
one can obtain the following reduced Hamiltonian by adiabatically
eliminating the degrees of freedom of the three-level artificial
atom~\cite{Villas-Boas and Almeida}:
\begin{equation}\label{Reduced Hamiltonian of the cavity mode}
H_c=\omega_c a^{\dagger}a +\xi (e^{-i(\tilde{\Omega}
t+\tilde{\phi}_0)}a^{\dagger2}+e^{i(\tilde{\Omega}
t+\tilde{\phi}_0)}a^2),
\end{equation}
where
\begin{eqnarray*}
\omega_c=\tilde{\omega}_c+\frac{2}{\delta}(|\lambda_g|^2+|\lambda_e|^2)
\end{eqnarray*}
is the effective frequency of the cavity mode; $\xi$ and
$\tilde{\phi}_0$ are {\it the effective amplitude and the initial
phase of the squeezed field}. The relation between $\xi$ and
$\tilde{\phi}_0$ is given by
\begin{eqnarray*}
\xi\exp(i\tilde{\phi}_0)=\frac{2}{\delta^2}\lambda_d\lambda_g\lambda_e.
\end{eqnarray*}
Notice that one can {\it continuously adjust $\xi$ by varying the
coupling strength $\lambda_d$ between the classical field and the
three-level artificial atom}.

\subsection{Tunable coupling between qubits}\label{s52}

In Fig.~\ref{Fig of the superconducting circuit}, let us now
consider the interaction between the two charge qubits and the
cavity field. After eliminating the degrees of freedom of the
auxiliary three-level system, we can obtain the following total
Hamiltonian of the charge qubits and the cavity field:
\begin{eqnarray}\label{Hamiltonian of the resonator and the two working qubits}
H&=&\xi(e^{-i(\tilde{\Omega} t+\tilde{\phi}_0)}a^{\dagger2}+e^{i(\tilde{\Omega} t+\tilde{\phi}_0)}a^2) \nonumber\\
&&+\sum_{j=1}^2 g(\eta_j-\cos\alpha_j
\sigma^{(j)}_{z}+\sin\alpha_j\sigma^{(j)}_{x})(a^{\dagger}+a)\nonumber\\
&&+\frac{1}{2}\sum_{j=1}^2\omega_{aj}\sigma^{(j)}_{z}+\omega_c
a^{\dagger}a,
\end{eqnarray}
where
\begin{eqnarray*}
\omega_{aj}=\sqrt{E_J^2(\Phi_{xj})+E_C^2(n_{gj})}
\end{eqnarray*}
has the same meaning with the corresponding quantity in
Sec.~\ref{s3} and \ref{s4}; $g=-e(C_g/C_{\Sigma})V_{{\rm rms}}^0$
is the coupling strength between the resonator and a single qubit;
$C_{\Sigma}$ is the total capacitance of a qubit; $V_{{\rm
rms}}^0=\sqrt{\omega_c/2C_r}$ is the root mean square (rms) of the
voltage across the $LC$ circuit; $C_r$ is the capacitance of the
resonator; $\eta_j=1-2n_{gj}$; and the angle
$\alpha_j=\arctan[E_J(\Phi_{xj})/E_C(1-2n_{gj})]$.


By using the rotating-wave approximation and assuming that
$n_{gj}=1/2\,(j=1,\,2)$, the total Hamiltonian $H$ in
Eq.~(\ref{Hamiltonian of the resonator and the two working
qubits}) can be rewritten as~\cite{Walls}:
\begin{eqnarray*}
H_{JC}&=&\omega_c
a^{\dagger}a+\sum_{j=1}^2\frac{E_J(\Phi_{x})}{2}\sigma^{(j)}_{z}+\sum_{j=1}^2
g(a^{\dagger}\sigma^{(j)}_{-}+a\sigma^{(j)}_{+}),\nonumber\\
&&+\xi (e^{-i(\tilde{\Omega}
t+\tilde{\phi}_0)}a^{\dagger2}+e^{i(\tilde{\Omega}
t+\tilde{\phi}_0)}a^2).
\end{eqnarray*}
Here, to simplify our discussions, we have set
$\Phi_{x1}=\Phi_{x2}=\Phi_x$.

We now assume that the qubits and the cavity field are in the
dispersive regime, i.e.,
\begin{eqnarray*}
\Delta=[E_J(\Phi_x)-\omega_c]\sim 100\;{\rm MHz}\gg|g|\sim10\;{\rm
MHz}.
\end{eqnarray*}
Thus, we can introduce the following unitary transformation to
diagonalize the Hamiltonian
$H_{JC}$:
\begin{eqnarray*}
U=\exp\left[\frac{g}{\Delta}\sum_{j=1}^2(a\sigma^{(j)}_{+}-a^{\dagger}\sigma^{(j)}_{-})\right].
\end{eqnarray*}
Up to first order in $g/\Delta$, we have
\begin{eqnarray*}
&&UH_{JC}U^{\dagger}
\approx\omega_c a^{\dagger}a+\xi e^{-i(\tilde{\Omega}t+\tilde{\phi}_0)}a^{\dagger 2}+\xi e^{i(\tilde{\Omega}t+\tilde{\phi}_0)}a^2\\
&&+\sum_{j=1}^2\left[\frac{\tilde{\omega}_a}{2}+\frac{4g^2}{\Delta^2}(\xi e^{-i(\tilde{\Omega}t+\tilde{\phi}_0)}a^{\dagger 2}+{\rm h.c.})+\frac{4g^2}{\Delta}a^{\dagger}a\right]\sigma^{(j)}_{z}\\
&&+\sum_{j=1}^2\left[\left(\frac{2g\xi
e^{-i(\tilde{\Omega}t+\tilde{\phi}_0)}}{\Delta}a^{\dagger}+\frac{g^2\xi
e^{-i(\tilde{\Omega}t+\tilde{\phi}_0)}}{\Delta^2}\right)\sigma^{(j)}_{+}+{\rm h.c.}\right]\\
&&+\mu_1(e^{-i(\tilde{\Omega}t+\tilde{\phi}_0)}\sigma^{(1)}_{+}\sigma^{(2)}_{+}+e^{i(\tilde{\Omega}t+\tilde{\phi}_0)}\sigma^{(1)}_{-}\sigma^{(2)}_{-})\\
&&+\mu_2(\sigma^{(1)}_{+}\sigma^{(2)}_{-}+\sigma^{(1)}_{-}\sigma^{(2)}_{+}),
\end{eqnarray*}
where
\begin{eqnarray*}
&\tilde{\omega}_a=E_J(\Phi_x)+4g^2/\Delta,& \\
&\mu_1=2g^2\xi/\Delta^2,\quad\mu_2=g^2/\Delta.&
\end{eqnarray*}

By adiabatically eliminating the degrees of freedom of the
resonator, the following two-qubit Hamiltonian can be obtained:
\begin{eqnarray*}
H_A&\approx&\sum_{j=1}^2\frac{E_J(\Phi_x)}{2}\sigma^{(j)}_{z}+\mu_2(\sigma^{(1)}_{+}\sigma^{(2)}_{-}+\sigma^{(1)}_{-}\sigma^{(2)}_{+})\\
&&+\mu_1(e^{-i(\tilde{\Omega}
t+\tilde{\phi}_0)}\sigma^{(1)}_{+}\sigma^{(2)}_{+}+e^{i(\tilde{\Omega}
t+\tilde{\phi}_0)}\sigma^{(1)}_{-}\sigma^{(2)}_{-}).
\end{eqnarray*}
Here, we have omitted all the single-qubit terms induced by the
interaction between qubits and the resonator because of the
conditions:
\begin{eqnarray*}
E_J(\Phi_x)/2\gg g^2/\Delta,\,\,\xi g/\Delta.
\end{eqnarray*}
As analyzed in subsection A of section~\ref{s5}, we can
continuously adjust the parameter $\xi$, thus the coupling
strength $\mu_1$ is continuously tunable.

Since the two superconducting charge qubits also interact with the
uncontrollable degrees of freedom in the environment (e.g.,
quantum noises induced by charge fluctuations on the electric
gates), the discussed two-qubit system should be considered as an
open quantum system. For this two-qubit system, the master
equation (\ref{General master equation}) can be obtained under the
Born-Markov approximation~\cite{Blais2}. 

From Eq.~(\ref{Optimal condition for weak interaction regime}), at
the charge degenerate points for both qubits, we know that the
optimal concurrence $C_{{\rm max}}\approx 0.31$ and fidelity
$F_{{\rm max}}\approx 0.65$ can be obtained when
\begin{eqnarray*}
\tilde{\Omega}=2E_J(\Phi_x),\quad\xi=\frac{1}{\sqrt{5}+1}\times\frac{\Delta^2}{g^2}\Gamma_1.
\end{eqnarray*}
Using now the same experimental parameters from
Ref.~\cite{Blais2}:
\begin{eqnarray*}
&\Delta=E_J-\omega_r=5\;{\rm GHz}-4.8\;{\rm
GHz}=200\;{\rm MHz},&\\
&g=20\;{\rm MHz},\,\,\Gamma_1/2\pi\sim0.1\;{\rm MHz},&\\
&\lambda_g,\,\lambda_e\sim10\;{\rm MHz},\,\,\delta\sim100\;{\rm
MHz},&
\end{eqnarray*}
the squeezed amplitude $\xi$ is of the order of $1$ MHz, which can
be realized by a strong microwave driving field with coupling
strength $\lambda_d\sim 100$ MHz. These parameters show that our
entanglement-protection proposal is experimentally realizable.

Although this section mainly concentrates on how to protect
entanglement in two charge qubits coupled to a transmission line
resonator, our proposal is also extendable to two flux qubits in a
coplanar transmission line resonator~\cite{Lindstrom}. Since the
system parameters~\cite{Lindstrom} are almost of the same order of
those for charge qubits, there is no essential difference between
them, as far as applying our proposal.

\section{Conclusions}\label{s6}

In summary, we propose a strategy to protect quantum entanglement
against independent quantum noises in several types of
superconducting circuits. For two superconducting qubits coupled
strongly via an inductive or a capacitive element, one can tune
the stationary entanglement by varying the single-qubit
oscillating frequencies or the coupling strengths between the
qubits. For superconducting qubits weakly coupled via a quantum
cavity (e.g., a transmission line resonator), an auxiliary
superconducting three-level system~\cite{Liu3,You4} with
$\Delta$-shaped transition is introduced to induce a controllable
squeezed field in the cavity. Such a controllable quantum squeezed
field can be further used to entangle two qubits in open
environments. Optimally, one can obtain a maximum concurrence of
about $0.31$ and a maximum fidelity of about $0.65$ for the above
systems.

Even though the proposed strategy can be used to protect quantum
entanglement, the obtained entanglement may not be high enough to
be used in quantum information processing. Additional entanglement
purification processes (e.g.,~\cite{Maruyama and Tanamoto}) should
be introduced to increase the stationary entanglement. These
procedures could make the superconducting circuit too complex. For
this reason, further research (possibly using the methods in
Ref.~\cite{MZhang}) will be focused on modifying our proposal to
obtain higher stationary entanglement.

Another interesting problem would be to develop a short-time
regime entanglement-protection strategy, e.g., to protect
entanglement during the gate operation process. In this regime,
the correlation effects of the environmental noises should be
considered, which leads to non-Markovian noises~\cite{Strunz and
Yu}. Different effects may be produced under non-Markovian noises.
The existing decoherence suppression strategies~\cite{Gordon and
Viola,WCui} against non-Markovian noises may be helpful to solve
this problem.
\\[0.2cm]

\begin{center}
\textbf{ACKNOWLEDGMENTS}
\end{center}

FN acknowledges partial support from the National Security Agency
(NSA), Laboratory Physical Science (LPS), Army Research Office
(ARO), National Science Foundation (NSF) grant No. EIA-0130383,
JSPS-RFBR 06-02-91200. J. Zhang was supported by the National
Natural Science Foundation of China under Grant Nos. 60704017,
60433050, 60635040, 60674039 and China Postdoctoral Science
Foundation. T. J. Tarn would also like to acknowledge partial
support from the U.S. Army Research Office under Grant
W911NF-04-1-0386.
\\[0.2cm]

\appendix
\section{Derivation of the maximum concurrence and
fidelity}\label{Derivation of the maximum concurrence and
fidelity}
In this appendix, we show the derivation of the
concurrence $C$ of the stationary state $\rho_{\infty}$ and the
fidelity $F$ between $\rho_{\infty}$ and the maximally-entangled
state $\rho_m$. Thus, we will derive Eq.~(\ref{Stationary
concurrence and fidelity for strong interaction regime}) in the
main text.

In order to simplify our discussions, let us use the so-called
coherent vector picture as in Refs.~\cite{Alicki,Altafini and
Zhang}. Considering the inner product $\langle X, Y\rangle={\rm
tr}(X^{\dagger}Y)$, we can find the following matrix basis for all
two-qubit matrices:
\begin{eqnarray}\label{Matrix basis for two-qubit operators}
&\left\{\frac{1}{2}I_{4 \times
4},\,\Omega_{14}^x,\,\Omega_{14}^y,\,\Omega_{23}^x,\,\Omega_{23}^y,\,\frac{1}{2}\sigma_{x}^{(1)},\,\frac{1}{2}\sigma_{y}^{(1)},\right.\nonumber&\\
&\frac{1}{2}\sigma_{x}^{(2)},\frac{1}{2}\sigma_{y}^{(2)},\,\frac{1}{2}\sigma_{x}^{(1)}\sigma_{z}^{(2)},\,\frac{1}{2}\sigma_{z}^{(1)}\sigma_{x}^{(2)},\,\frac{1}{2}\sigma_{y}^{(1)}\sigma_{z}^{(2)},\nonumber&\\
&\left.\frac{1}{2}\sigma_{z}^{(1)}\sigma_{y}^{(2)},\,\Omega_{14}^z,\Omega_{23}^z,\frac{1}{2}\sigma_{z}^{(1)}\sigma_{z}^{(2)}\right\},&
\end{eqnarray}
where $I_{4\times 4}$ is the $4\times 4$ identity matrix, and
$\Omega_{14}^x$, $\Omega_{14}^y$, $\Omega_{23}^x$,
$\Omega_{23}^y$, $\Omega_{14}^z$, $\Omega_{23}^z$ are defined as:
\begin{eqnarray*}
\Omega_{14}^x=\left(%
\begin{array}{cccc}
   &  &  & \frac{1}{\sqrt{2}} \\
   &  & 0 &  \\
   & 0 &  &  \\
  \frac{1}{\sqrt{2}} &  &  &  \\
\end{array}%
\right),\,\,\Omega_{14}^y=\left(%
\begin{array}{cccc}
   &  &  & \frac{-i}{\sqrt{2}} \\
   &  & 0 &  \\
   & 0 &  &  \\
  \frac{i}{\sqrt{2}} &  &  &  \\
\end{array}%
\right),\\
\Omega_{23}^x=\left(%
\begin{array}{cccc}
   &  &  & 0 \\
   &  & \frac{1}{\sqrt{2}} &  \\
   & \frac{1}{\sqrt{2}} &  &  \\
  0 &  &  &  \\
\end{array}%
\right),\,\,\Omega_{23}^y=\left(%
\begin{array}{cccc}
   &  &  & 0 \\
   &  & \frac{-i}{\sqrt{2}} &  \\
   & \frac{i}{\sqrt{2}} &  &  \\
  0 &  &  &  \\
\end{array}%
\right),\\
\Omega_{14}^z=\left(%
\begin{array}{cccc}
  \frac{1}{\sqrt{2}} &  &  &  \\
   & 0 &  &  \\
   &  & 0 &  \\
   &  &  & \frac{-1}{\sqrt{2}} \\
\end{array}%
\right),\,\,\Omega_{23}^z=\left(%
\begin{array}{cccc}
  0 &  &  &  \\
   & \frac{1}{\sqrt{2}} &  &  \\
   &  & \frac{-1}{\sqrt{2}} &  \\
   &  &  & 0 \\
\end{array}%
\right).
\end{eqnarray*}
The system density matrix $\rho$ can be expanded under this matrix
basis as:
\begin{eqnarray*}
\rho=\frac{1}{4}I_{4\times 4}+\sum_{i=1}^{15} m_i \Omega_i,
\end{eqnarray*}
where $\Omega_i \, (i=1,\cdots,15)$ are all traceless basis
matrices in (\ref{Matrix basis for two-qubit operators}) and
$m_i={\rm tr}(\Omega_i\rho)$.

Let $m=(m_1,\cdots,m_{15})^T$, and thus the master equation
(\ref{General master equation}) can be rewritten
as~\cite{Alicki,Altafini and Zhang}:
\begin{equation}\label{Coherent vector representation of the two-qubit equation}
\dot{m}=O_A\;m+D\;m+g,
\end{equation}
where $O_A$ is the adjoint representation matrix of $-iH_A$ and
($D m+g$) is the coherent vector representation of the Lindblad
terms:
\begin{eqnarray*}
\sum_{j=1}^2\Gamma_1\mathcal{D}[\sigma_{-j}]\rho+\sum_{j=1}^2
2\Gamma_{\phi}\mathcal[\sigma_{zj}]\rho,
\end{eqnarray*}
with $D\leq 0$ and $g$ a constant vector. Further, let
\begin{eqnarray*}
m^p&=&(m_{14}^x,m_{14}^y,m_{23}^x,m_{23}^y)^T,\\
m^{\eta}&=&(m_{14}^z,m_{23}^z,m_{zz})^T,\\
m^{\epsilon}&=&(m_{x0},m_{y0},m_{0x},m_{0y},m_{xz},m_{zx},m_{yz},m_{zy})^T,
\end{eqnarray*}
where
\begin{eqnarray*}
&m_{14}^{\alpha}={\rm
tr}(\Omega_{14}^{\alpha}\rho),\,m_{23}^{\beta}={\rm
tr}(\Omega_{23}^{\beta}\rho),\,\,\alpha,\beta=x,y,z,&\\
&m_{\alpha\beta}={\rm
tr}\left[\left(\frac{1}{2}\sigma^{(1)}_{\alpha}\sigma^{(2)}_{\beta}\right)\rho\right],\,\,\alpha,\beta=0,x,y,z,&
\end{eqnarray*}
and $\sigma^{(j)}_{0}=I_{2\times2}\,\,\,(j=1,\,2)$ are $2\times2$
identity matrices acting on the qubit $j$. Then, we can rewrite
Eq.~(\ref{Coherent vector representation of the two-qubit
equation}) as:
\begin{eqnarray}\label{Variable equation of the coherent vector representation of the two-qubit equation}
\dot{m}^p&=&O_0^p m^p+\sum_{i=1}^4 u_i O_i^{\eta} m^{\eta}+D^p m^p,\nonumber\\
\dot{m}^{\eta}&=&\sum_{i=1}^4 u_i(-O_i^{\eta\,T})m^p+D^{\eta}m^{\eta}+g^{\eta},\\
\dot{m}^{\epsilon}&=&\sum_{i=1}^4 u_i O_i^{\epsilon}
m^{\epsilon}+D^{\epsilon}m^{\epsilon},\nonumber
\end{eqnarray}
where $D^p=-4(\Gamma_1+2\Gamma_{\phi})I_{4\times 4}=-8\Gamma_2
I_{4\times 4}$ and
\begin{eqnarray*}
&u_1=8\mu_1\cos\theta_1,\,\,\,u_2=8\mu_1\sin\theta_1,&\\
&u_3=8\mu_2\cos\theta_2,\,\,\,u_4=-8\mu_2\sin\theta_2,&
\end{eqnarray*}
\begin{eqnarray*}
&O_0^p=\left(%
\begin{array}{cccc}
  0 & \Omega &  &  \\
  -\Omega & 0 &  &  \\
   &  & 0 & \omega_{a1}-\omega_{a2} \\
   &  & \omega_{a2}-\omega_{a1} & 0 \\
\end{array}%
\right),&
\end{eqnarray*}
\begin{eqnarray*}
&O_1^{\eta}=\left(%
\begin{array}{ccc}
  0 & 0 & 0 \\
  -1 & 0 & 0 \\
  0 & 0 & 0 \\
  0 & 0 & 0 \\
\end{array}%
\right),\,\,O_2^{\eta}=\left(%
\begin{array}{ccc}
  1 & 0 & 0 \\
  0 & 0 & 0 \\
  0 & 0 & 0 \\
  0 & 0 & 0 \\
\end{array}%
\right),
\end{eqnarray*}
\begin{eqnarray*}
&O_3^{\eta}=\left(%
\begin{array}{ccc}
  0 & 0 & 0 \\
  0 & 0 & 0 \\
  0 & 0 & 0 \\
  0 & -1 & 0 \\
\end{array}%
\right),\,\,O_4^{\eta}=\left(%
\begin{array}{ccc}
  0 & 0 & 0 \\
  0 & 0 & 0 \\
  0 & 1 & 0 \\
  0 & 0 & 0 \\
\end{array}%
\right),
\end{eqnarray*}
\begin{eqnarray}\label{Matrices in the coherent vector picture}
&D^{\eta}=\left(%
\begin{array}{ccc}
  -4\Gamma_1 & 0 & 0 \\
  0 & -4\Gamma_1 & 0 \\
  4\sqrt{2}\Gamma_1 & 0 & -8\Gamma_1 \\
\end{array}%
\right),\,\,g^{\eta}=\left(%
\begin{array}{c}
  2\sqrt{2}\Gamma_1 \\
  0 \\
  0 \\
\end{array}%
\right).&\nonumber\\
\end{eqnarray}
$D^{\epsilon}$ and $O_i^{\epsilon}$ in the last equation in
Eq.~(\ref{Variable equation of the coherent vector representation
of the two-qubit equation}) are respectively negative and
traceless skew-symmetric matrices.

With simple calculations, we can obtain the following stationary
solution of Eq.~(\ref{Variable equation of the coherent vector
representation of the two-qubit equation}):
\begin{eqnarray*}
&m^{\epsilon}(\infty)=0,\,\,m_{23}^x(\infty)=m_{23}^y(\infty)=m_{23}^z(\infty)=0,&\\
&m_{14}^x(\infty)=\frac{1}{\sqrt{2}}p\cos(\theta_1-\phi),\,\,m_{14}^y(\infty)=\frac{1}{\sqrt{2}}p\sin(\theta_1-\phi),&\\
&m_{14}^z(\infty)=\frac{\sqrt{2}}{4}\left\{1+\sqrt{1-\frac{8\Gamma_2}{\Gamma_1}p^2}\right\},&\\
&m_{zz}(\infty)=\frac{1}{4}\left\{1+\sqrt{1-\frac{8\Gamma_2}{\Gamma_1}p^2}\right\},&
\end{eqnarray*}
where $p$ and $\phi$ are given in Eq.~(\ref{p and phi for strong
interaction regime}), from which we can calculate the stationary
state $\rho_{\infty}$ and the stationary fidelity
$F(\rho_{\infty})$ in Eq.~(\ref{Stationary concurrence and
fidelity for strong interaction regime}).

Further, recall that the concurrence of the quantum state
\begin{eqnarray*}
\rho=\left(%
\begin{array}{cccc}
  a &  &  & w \\
   & b & z &  \\
   & z^* & c &  \\
  w^* &  &  & d \\
\end{array}%
\right)
\end{eqnarray*}
can be analytically solved as~\cite{Yu1}:
\begin{eqnarray*}
C(\rho)=2\max\{|w|-\sqrt{bc},|z|-\sqrt{ad},0\},
\end{eqnarray*}
from which we can obtain the stationary concurrence
$C(\rho_{\infty})$ in Eq.~(\ref{Stationary concurrence and
fidelity for strong interaction regime}).
\\[0.2cm]

\end{document}